
\input harvmac
\noblackbox
\advance\vsize 10pt

\input epsf
\ifx\epsfbox\UnDeFiNeD\message{(NO epsf.tex, FIGURES WILL BE
IGNORED)}
\def\figin#1{\vskip2in}
\else\message{(FIGURES WILL BE INCLUDED)}\def\figin#1{#1}\fi
\def\ifig#1#2#3{\xdef#1{fig.~\the\figno}
\goodbreak\topinsert\figin{\centerline{#3}}%
\smallskip\centerline{\vbox{\baselineskip12pt
\advance\hsize by -1truein\noindent{\bf Fig.~\the\figno:} #2}}
\bigskip\endinsert\global\advance\figno by1}

%
%
\def\frac#1#2{ {{#1} \over {#2}}}
\def\nv{\vec n}
\def\pa{\partial}
\def\pmin{p^-}
\def\pp{p^+}
\def\pv{\vec p}
\def\qm{q^-}
\def\qp{q^+}
\def\qv{\vec q}
\def\xm{x^-}
\def\xp{x^+}
\def\xv{\vec x}
\def\Xv{\vec X}
\def\yv{\vec y}

%
%
\def\CQG{{\it Class. Quan. Grav.\/}}
\def\PREP{{\it Phys. Rep.\/}}

\def\NP{{\it Nucl. Phys.\ }}

\def\PL{{\it Phys. Lett.\ }}
\def\PR{{\it Phys. Rev.\ }}
\def\PRL{{\it Phys. Rev. Lett.\ }}
\def\CMP{{\it Comm. Math. Phys.\ }}

%
%


\lref\stu{L.~Susskind, L.~Thorlacius, and J.~Uglum, \PR {\bf D48}
(1993) 3743.}

\lref\thooft{G.~'t~Hooft, \NP {\bf B256} (1985) 727; \NP {\bf B335}
(1990) 138; {\it Phys. Scr.} {\bf T36} (1991) 247, and references
therein.}

\lref\kvv{Y.~Kiem, E.~Verlinde, and H.~Verlinde, {\it Black Hole
Horizons and Complementarity,} preprint CERN-TH-7469/94, PUPT-1504,
hep-th/9502074.}

\lref\st{L.~Susskind and L.~Thorlacius, \PR {\bf D49} (1994) 966.}

\lref\comms{C.~R.~Stephens, G.~'t~Hooft, and B.~F.~Whiting, \CQG {\bf
11} (1994) 621; G.~'t~Hooft, {\it Horizon Operator Approach to Black
Hole Quantization,} preprint THU-94-02,
gr-qc/9402037.}

\lref\svv{E.~Verlinde and H.~Verlinde, \NP {\bf B406} (1993) 43;
K.~Schoutens, E.~Verlinde, and H.~Verlinde, \PR {\bf D48} (1993)
2690.}

\lref\hawking{S.~W.~Hawking, \CMP {\bf 43} (1975) 199.}

\lref\page{D.~N.~Page, \PRL {\bf 71} (1993) 1291.}

\lref\lenny{L.~Susskind, \PRL {\bf 71} (1993) 2367; \PR {\bf D49}
(1994) 6606.}

\lref\mpt{A.~Mezhlumian, A.~Peet, and L.~Thorlacius,
\PR {\bf D50} (1994) 2725.}

\lref\zuth{W.H.~Zurek and K.S.~Thorne, \PRL {\bf 54} (1985) 2171.}

\lref\lsju{L.~Susskind and J.~Uglum, \PR {\bf D50} (1994) 2700.}


\lref\wald{R.~M.~Wald, private communication at the Santa Barbara ITP
conference ``Quantum Aspects of Black Holes'', June 1993.}


\lref\lsu{D.~A.~Lowe, L.~Susskind, and J.~Uglum, \PL {\bf B327}
(1994) 226.}

\lref\atwit{J.~J.~Atick and E.~Witten, \NP {\bf B310} (1988) 291.}

\lref\klebsus{I.~Klebanov and L.~Susskind, \NP {\bf B309} (1988)
175.}

\lref\gross{D.~J.~Gross, \PRL {\bf 60} (1988) 1229.}

\lref\holo{L.~Susskind, {\it The World as a Hologram}, preprint
SU-ITP-94-33, hepth/\-9409089.}


\lref\mand{S.~Mandelstam, \NP {\bf B64} (1973) 205; {\bf B83} (1974)
413; \PREP {\bf 13} (1974) 259.}

\lref\kaki{M.~Kaku and K.~Kikkawa, \PR {\bf D10} (1974) 1110, 1823.}

\lref\crge{E.~Cremmer and J.-L.~Gervais, \NP {\bf B76} (1974) 209;
{\bf B90} (1975) 410.}

\lref\htc{J.~F.~L.~Hopkinson, R.~W.~Tucker, and P.~A.~Collins,
\PR {\bf D12} (1975) 1653.}

\lref\martinec{E.~Martinec, \CQG {\bf 10} (1993) L187;
{\it Strings and Causality,} preprint EFI-93-65, hep-th/9311129.}

\lref\lowe{D.~A.~Lowe, \PL {\bf B326} (1994) 223.}

\lref\bd{J.~D.~Bjorken and S.~D.~Drell, {\it Relativistic Quantum
Fields,} McGraw Hill, 1965.}

\lref\gsw{M.~B.~Green, J.~H.~Schwarz, and E.~Witten, {\it Superstring
Theory, Vol. 2}, Cambridge University Press, 1987.}

%
%
\Title{\vbox{\baselineskip12pt \hbox{hep-th/9506138}
\hbox{NSF-ITP-95-47} \hbox{UCSBTH-95-12} \hbox{SU-ITP-95-13}}}
{\vbox{
\hbox{\centerline{Black Hole Complementarity {\it vs.} Locality}}}}
\centerline{David A. Lowe,${}^a$ Joseph Polchinski,${}^b$
Leonard Susskind,${}^c$}
\centerline{L\'arus Thorlacius,${}^b$ and John Uglum${}^c$}
\vskip 1.2cm
\vbox{%
\hbox to\the\hsize{\vbox{%
\hbox{\sl a) Department of Physics\hfil}
\hbox{\sl University of California\hfil}
\hbox{\sl Santa Barbara\hfil}
\hbox{\sl CA 93106-4030\hfil}
}\hfil
\vbox{%
\hbox{\sl b) Inst.~for Theoretical Physics\hfil}
\hbox{\sl University of California\hfil}
\hbox{\sl Santa Barbara\hfil}
\hbox{\sl CA 93106-4030\hfil}
}\hfil
\vbox{%
\hbox{\sl c) Department of Physics\hfil}
\hbox{\sl Stanford University\hfil}
\hbox{\sl Stanford\hfil}
\hbox{\sl CA  94305-4060\hfil}
}}}
\vskip 1cm
\noindent
{\baselineskip 14pt
The evaporation of a large mass black hole can be described
throughout most of its lifetime by a low-energy effective
theory defined on a suitably chosen set of smooth spacelike
hypersurfaces.  The conventional argument for information
loss rests on the assumption that the effective theory is
a {\it local\/} quantum field theory.  We present evidence that this
assumption fails in the context of string theory.  The commutator of
operators in light-front string theory, corresponding to certain
low-energy observers on opposite sides of the event horizon,
remains large even when these observers are spacelike
separated by a macroscopic distance.  This suggests that
degrees of freedom inside a black hole should not be viewed
as independent from those outside the event horizon.  These
nonlocal effects are only significant under extreme kinematic
circumstances, such as in the high-redshift geometry of a
black hole.  Commutators of space-like separated operators
corresponding to ordinary low-energy observers in Minkowski
space are strongly suppressed in string theory.
}

\Date{June, 1995}

\newsec{Introduction}

According to the principle of black hole
complementarity~\refs{\stu}, an observer who remains outside
the horizon of a black hole can describe the black hole as a very
hot membrane, the stretched horizon, which lies
just above the mathematical event horizon, and absorbs any
matter, energy, and information which fall onto it.  The information
which is absorbed by the stretched horizon is eventually re-emitted
in the Hawking radiation, albeit in a very
scrambled form.  An observer who falls freely into the black hole
sees things very differently: no membrane, no high temperature, no
irregularities of any kind as the observer crosses the event horizon.
 The principle of black hole complementarity asserts the consistency
of these apparently contradictory descriptions.  According to this
principle, the matter which has fallen past the event horizon and the
Hawking radiation are not different objects.  They are complementary
descriptions of a single system, viewed from very different reference
frames which are related by an enormous Lorentz boost.  A similar
viewpoint has long been advocated by 't~Hooft~\refs{\thooft} and more
recently by Kiem, Verlinde, and Verlinde~\refs{\kvv}.
At present, such strange behavior cannot be
ruled out because it involves physics at energy scales far beyond
anything with which we have any experience~\refs{\st}.

Although no logical contradiction is known to follow from the
principle of black hole complementarity, it nevertheless seems to
contradict our ordinary ideas about locality.
In actuality, black hole complementarity does not require any
observer
to detect nonlocal effects.  In the membrane picture the
observation of information in the Hawking radiation by a distant
observer involves nothing acausal, since the Hawking radiation is in
causal contact with the stretched horizon at all times.
As for infalling observers, they see the ordinary
low-energy laws of nature until the singularity is approached.  It is
only in certain correlations between events on either
side of the horizon that nonlocality is required.  Such correlations
are unobservable in the sense that they cannot be established by any
single observer without violating known laws of physics~\refs{\st}.
On the other hand, if physics is described in terms of a local
effective field theory, then correlations across the event horizon
will in fact be in contradiction with black hole complementarity,
as we will discuss below.  If black hole complementarity is correct,
the usual principles of local quantum field theory must break
down, not only at small scales, but at all scales.  And yet these
violations of locality must be undetectable in ordinary low-energy
experiments.  Their only role should be to reconcile the two
complementary descriptions.  Evidently, the nonlocality must be of
an extraordinarily special and subtle kind.

The nature of the required nonlocality can be illustrated by
examining an argument, which we will call the {\it nice-slice
argument,} which is often put forward in support of the idea of
information loss.  The essence of this argument is very simple.
Since the process of gravitational collapse and subsequent
evaporation of a very large black hole begins and ends with very
low-energy particles, and since the evolution of the black hole is
very
slow on microscopic time scales, the adiabatic theorem should ensure
that high-energy degrees of freedom decouple.  In other words, the
process can be described using only a local low-energy effective
field theory defined on a (slowly varying) background geometry, and
fluctuations of the gravitational field can be neglected.  Of course,
the final burst of energy involves a few high-energy particles,
but this is irrelevant for our considerations, since such a small
number of particles can not carry off an appreciable amount of the
information that originally fell into the black hole.  It is then
straightforward to argue that the known behavior of local field
theories prohibits information retrieval.

The nice-slice argument is formalized in Section 2 of this paper.  In
Section 3, we present evidence that string theory fails to meet one
very important assumption of the nice-slice argument -- the
assumption of locality in the low-energy nice-slice theory.  To do
so, we construct a low-energy nice-slice theory using light-front
string field theory, and calculate the commutators of low-energy
nice-slice fields.  It turns out that commutators of nice-slice
fields behind the horizon do not commute with nice-slice fields in
front of the horizon.  The idea that operators behind the horizon
fail to commute with operators in front of the horizon has been
advocated previously by several authors~\refs{\thooft, \kvv, \comms,
\svv}.
The conclusion is that there must exist nonlocal states in the
low-energy nice-slice theory.  It will be shown that to leading order
in the string coupling, these states are highly excited strings
stretched between points on the nice slice.

We present most of our calculations in the two Appendices.  This
allows us to better focus on ideas and results in the main text.
Our calculations are strictly
speaking only valid in the limit of very large black hole mass which
allows us to neglect the effects of the local curvature in the region
of interest.  We also only work to leading order in string
perturbation theory which further restricts the range of validity of
the calculations.  In spite of these
technical limitations, our results demonstrate a basic difference
between light-front string theory and quantum field theory.

Before proceeding, it is important to discuss the various scales of
size and energy which occur in the discussion of black hole
evaporation.  The largest energies, which we will call {\it
trans-Planckian,} are vastly larger than the Planck mass $M_P$.  For
a black hole of mass $M$, the trans-Planckian energies involved can
be of order $M_P \exp (G M^2)$.  Recall that according to the
conventional analysis of Hawking radiation~\refs{\hawking}, the
outgoing radiation originates in incoming vacuum fluctuations which
become deformed by the black hole geometry.  After a time of order
$G^2 M^3$, an appreciable fraction of the black hole energy has been
radiated away, and information is expected to appear in the
evaporation products~\refs{\page}.  At this stage, the outgoing
Hawking radiation is associated with infalling modes with
trans-Planckian energies of order $M_P \exp (G M^2)$.

It is generally felt, however, that a correct understanding of
Hawking evaporation should not require knowledge of trans-Planckian
physics.  According to black hole complementarity, this is correct
for
observers who remain strictly outside the horizon.  For such
observers, a description in terms of a stretched horizon composed of
Planck scale degrees of freedom should suffice.  Using coordinates
which
lie partly behind the horizon, however, we will find that
trans-Planckian
degrees of freedom are important.  In particular, an understanding of
the redundancy of degrees of freedom on either side of the horizon
requires a correct treatment of these extremely high-energy degrees
of freedom.  In free field
theory, the trans-Planckian modes are associated with distance and
time scales of order $\ell_P \exp (-G M^2)$.  In the nice-slice
argument, it is not necessary to assume that free-field theory is
valid
for the trans-Planckian modes, but it is assumed that they can be
localized on some scale small compared to the overall geometry, such
as
$\ell_P$.  We shall see that in
string theory, the relevant trans-Planckian modes correspond to
distances of order $\ell_P \exp (G M^2)$.  The nonlocality induced
by trans-Planckian modes can thus extend to very large distances.

In the membrane picture the degrees of freedom contained in the
region of order the Planck or string scale from the event horizon
comprise the stretched horizon.  These degrees of freedom store and
thermalize information from the viewpoint of the external
observer~\refs{\lenny,\mpt}, and they are responsible for the
entropy of the black hole~\refs{\thooft, \zuth, \lsju}.
Finally, there are low-energy modes well below the Planck or string
scale, which correspond to the energies of Hawking particles that
escape from the black hole.  The nice-slice argument suggests
that only these low-energy modes are
essential to a complete understanding of Hawking evaporation.
We will argue that this is not the case in string theory.

\newsec{The Nice-Slice Argument}

Consider the process by which a collection of particles of total mass
$M$ gravitationally collapses to form a black hole.   In the absence
of any net angular momentum, the black hole becomes approximately
spherically symmetric in a short time after the collapse process.
As long as we are only considering
processes with lifetimes short compared to $G^2 M^3$, we can
approximate the geometry of the system by the Schwarzschild solution.
 The nice-slice argument begins by introducing a family of Cauchy
surfaces which foliate the geometry.  The surfaces should
avoid regions of strong spacetime curvature and yet cut through
the infalling matter and the outgoing Hawking radiation so that both
sets of particles have low energy in the local frame of the slice.
Also, in order to ensure that short distance physics does not
creep in to the description through the choice of coordinates, we
require that the slices be everywhere smooth, with small extrinsic
curvature compared to any microscopic scale.  For convenience, we
will choose surfaces that agree with surfaces of constant
Schwarzschild time far from the black hole.  Such a family of
surfaces
will henceforth be designated ``nice slices''.  While it is seldom
spelled out, the existence of such a set of surfaces is implicitly
assumed in much of the existing literature on black hole evaporation.
The first explicit construction of nice slices that we are aware of
was carried out by Wald~\refs{\wald}.

\ifig\fone{Kruskal diagram of static black hole solution.}
{\epsfxsize=5.5cm \epsfysize=5.5cm \epsfbox{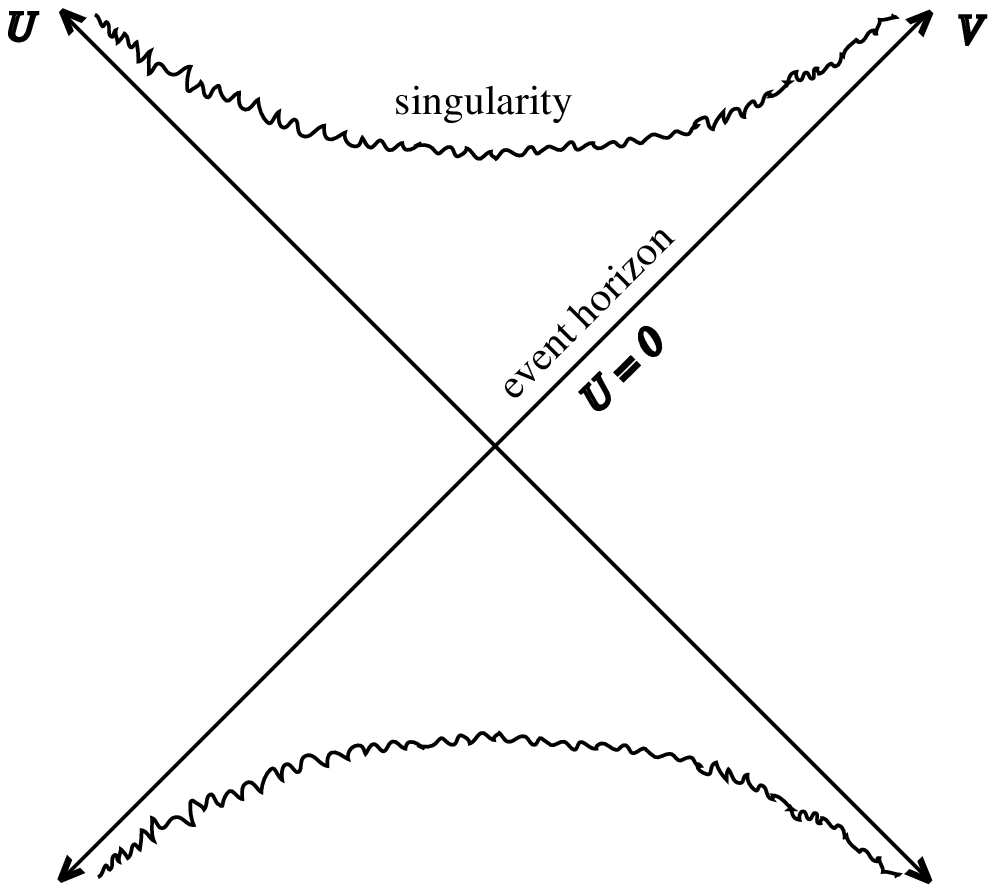}}

To construct an example of a family of nice slices, we begin with the
Schwarzschild
black hole in Kruskal-Szekeres coordinates, as shown in \fone.
These coordinates are related to the usual Schwarzschild coordinates
$r$ and $t$ by the transformation
\eqn\kruskals{
\eqalign{
-UV &= 16G^2 M^2 \left ( {r \over {2GM}} - 1 \right ) e^{r/2GM}\>,
\cr
-V/U &= e^{t/2GM} \>. \cr
}
}
The singularity is at $UV = 16G^2 M^2$, and the event
horizon is the surface $U=0$.

We first construct a spacelike surface composed of two pieces joined
smoothly at the surface $U=V$.  The first piece is the hyperbola
which satisfies the condition $UV = R^2$ for $V < U$.  The constant
$R$ is assumed to be large by comparison with any microscopic scale,
but should not be so large that the surface is anywhere near the
singularity.  Later, we will choose $R$ to be fixed and send
$M \rightarrow \infty$.  The second piece of the nice slice is
defined by the line satisfying $U+V = 2R$ for $V > U$.   The
resulting surface is asymptotic to the surface defined by $t = 0$.
\ifig\ftwo{A family of nice slices.}
{\epsfxsize=5.5cm \epsfysize=5.5cm \epsfbox{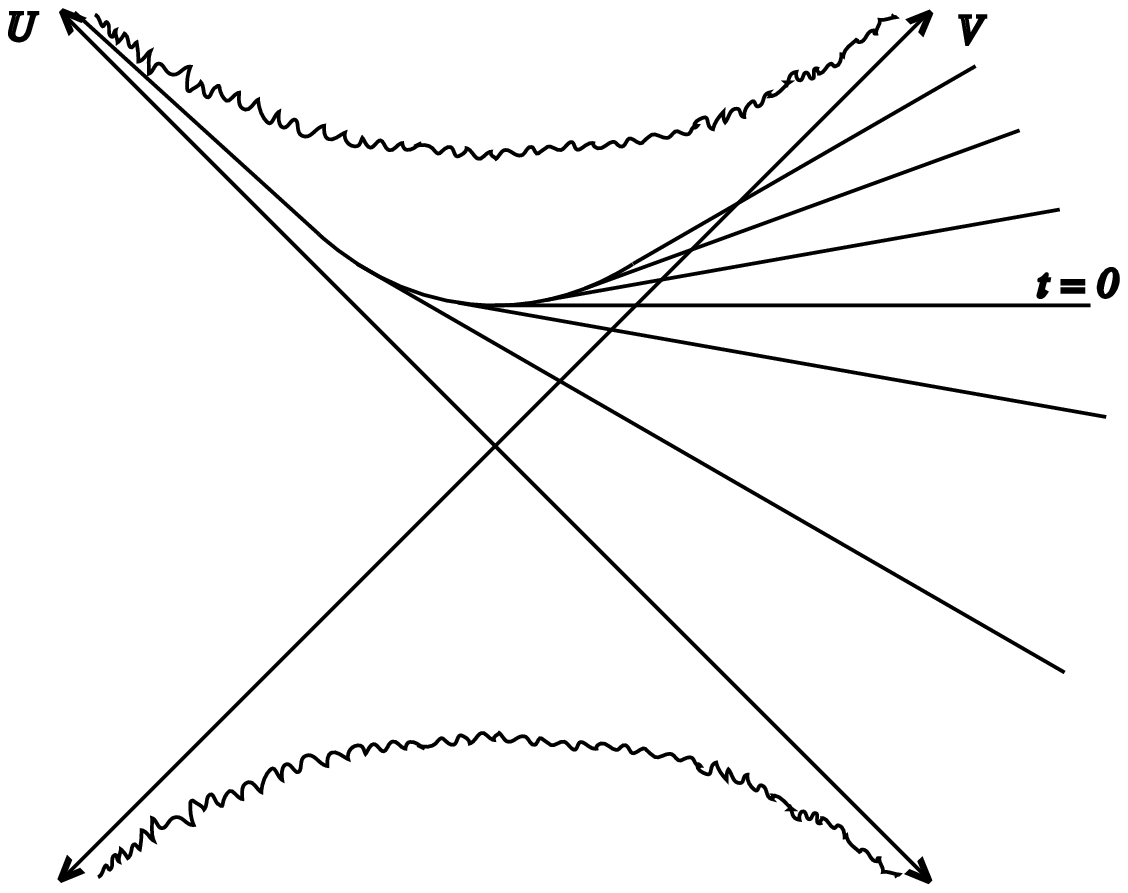}}

The nice slice we have constructed can be pushed forward and backward
in time by using the symmetry under the boost-like operations
\eqn\boost{
\eqalign{
U &\rightarrow U' = U e^{-t/4GM} \>,\cr
V &\rightarrow V' = V e^{t/4GM} \>. \cr
}
}
Since the nice slices are asymptotic to surfaces of constant
Schwarzschild time, they can be parametrized by $t$.  The full set of
nice slices can then be written
\eqn\niceslices{
\eqalign{
UV &= R^2 \>, \qquad V < e^{t/2GM} U \>, \cr
e^{t/4GM} U + e^{-t/4GM} V &= 2R \>, \qquad V > e^{t/2GM} U \>, \cr
}
}
and is shown in \ftwo.  The join between the line segment and the
hyperbola on each slice should be smoothed to avoid having a large
extrinsic curvature there.

A Hamiltonian may be defined by introducing a vector field $v$ which
is orthogonal to the nice slices.  The nice-slice Hamiltonian,
$H_{NS}$, is the generator of motions along this vector field, and
maps the state of the system on one slice to a state on a
neighboring slice by means of the Schr\"odinger equation
\eqn\schrodinger{
i \pa_t | \psi \rangle = H_{NS} | \psi \rangle \>.
}
Since the vector field $v$ is not a Killing vector field, the
Hamiltonian $H_{NS}$ is time dependent, so there is no conserved nice
slice energy.  If the mass $M$ of the black hole and the value of $R$
are very large, however, the rate of change of the nice-slice energy
is very small, and becomes adiabatic as $M$ and $R$ tend to infinity.

The three-momentum of an infalling particle as it crosses a given
slice can be defined by projecting the four-momentum vector onto the
slice.  It is easily seen that the spatial momentum of an ordinary
particle remains small throughout its entire journey toward the
hyperbola $UV = R^2$.  This is in sharp contrast to the situation in
Schwarzschild coordinates, where the three-momentum of a particle
diverges as it approaches the horizon.  The outgoing Hawking quanta
also have low momenta as they cross each nice slice.

So far we have been discussing a classical background geometry,
but in order to address the issue of information loss we should
include the effect of black hole evaporation.  This introduces an
important new feature into the problem.  Our nice slices in the
static classical geometry never get closer to the singularity than
the hyperbola $UV=R^2$, but when the semi-classical back-reaction
is included the black hole has a finite lifetime and eventually
the part of a slice that extends into the black hole interior will
encounter large curvature.  The best one can do in this case is
to construct a family of slices that avoid the region of strong
gravitational coupling for as long as possible.  This is not a
serious drawback.  If we start with a sufficiently large black
hole then even after 99.99\% of the energy has evaporated the
black hole will still be large and a set of nice slices can still
be found.  By starting with a sufficiently large black hole, we can
thus follow the evolution long enough to have most of the total
Hawking radiation already emitted, and well separated from the
black hole region, by the time our nice slices run into strong
curvature.

It is therefore natural to make the assumption that the
entire history of the black hole (except for the short time when the
Hawking temperature
exceeds the cutoff scale) can be described by a low-energy effective
quantum field theory defined on a slowly varying background
geometry.  The cutoff length is chosen sufficiently large that
gravitational fluctuations can be ignored altogether.
We emphasize that no trans-Planckian frequencies have entered into
the discussion.  In fact the adiabatic theorem implies that only
degrees of freedom of very low-energy ($E\sim 1/M$) get excited
from their ground state in the Hawking emission process.  The
construction seems to leave no room for stringy short distance
behavior, or any other modification of field theory, to influence the
details of the Hawking radiation.

The nice-slice argument can be combined with another argument, the
``no quantum Xerox principle''~\refs{\stu}, to show that the
information carried by infalling matter can not be stored on the
stretched horizon.  Consider a late time slice $\Sigma$, and
partition it into two portions, $\Sigma_{in}$ and $\Sigma_{out}$,
with $\Sigma_{in}$ containing the region behind the horizon,
while $\Sigma_{out}$ contains the horizon and the region outside
the black hole.  Since the low-energy nice-slice theory is a local
quantum field theory on a time-dependent but classical background
geometry, low-energy field operators will commute when their
arguments are spacelike related (as measured by the background
metric).  This means that we can form a complete set of commuting
observables using local fields defined on $\Sigma_{in}$ and
$\Sigma_{out}$, and thus the Hilbert space of states on the slice
$\Sigma$ factorizes into a tensor product
\eqn\tensor{
{\cal H}(\Sigma) = {\cal H}(\Sigma_{in}) \otimes {\cal
H}(\Sigma_{out}) \>.
}
The evolution operator defines a linear map from states in the
Hilbert space of initial configurations of infalling matter
to states in ${\cal H}(\Sigma)$.

It is now easy to give the argument for information loss.  The ``no
quantum Xerox principle'' states that the process of linear evolution
cannot faithfully replicate quantum information in two separate sets
of commuting degrees of freedom.  If the infalling information is
completely and faithfully recorded in the states of ${\cal
H}(\Sigma_{in})$ (as is widely believed), then little or no
information can be found in the states of ${\cal
H}(\Sigma_{out})$.  This conclusion would apply to both the Hawking
radiation itself and to the stretched horizon.

\newsec{The Nice Slice and String Field Theory}

Although the nice-slice argument seems very general, we will present
evidence from string theory that it fails. A key
assumption that goes into the nice-slice argument is that
the underlying microscopic theory is
approximately a local field theory.  Specifically, it assumes that
any nonlocality which results from the non-zero size of strings is
limited to small space-like separations, but this is incorrect, as we
shall see later on.

The analysis begins by considering two spacetime points, $x_1$ and
$x_2$, which lie on a fixed nice slice $\Sigma_t$ corresponding to
Schwarzschild time $t$.  The time $t$ is chosen large, but not so
large that an appreciable amount of evaporation has occurred.  The
point $x_2$ lies behind the horizon, and could be chosen to lie on
the hyperbola $UV=R^2$.  It may be thought of as a point near the
trajectory of a low-energy particle which has fallen through the
horizon at some early time.  Point $x_1$ lies outside the event
horizon and should be associated with the stretched horizon.
According to the principle of black hole complementarity, an
observer who remains permanently outside the black
hole sees the
infalling information stored in this region.  The point $x_1$ may be
chosen according to the following
procedure.
First, consider the nice slice $\Sigma_0$, and pick a point outside
the event horizon, for example, the point $U = U_0 < 0$, $V = V_0 >
0$.  Now push the point forward in time along a timelike Killing
vector until it arrives at the point
\eqn\newpoint{
\eqalign{
U_1 &= U_0 e^{-t/4GM} \cr
V_1 &= V_0 e^{t/4GM} \>, \cr
}
}
which lies on $\Sigma_t$.  As $t$ increases, the spacelike separation
between $x_1$ and $x_2$ grows like $\exp (t/4GM)$.

We will assume that some form of string field theory allows us to
obtain component fields $\phi_A(x)$ for each mass eigenstate of the
string.  A low-energy effective field ${\hat \phi}_A(y)$ can then be
defined by
\eqn\lefdef{
{\hat \phi}_A(y) = \int d^D x f(x-y) \phi_A (x) \>,
}
where the test function $f$ is assumed to be smooth enough to
eliminate variations of $\phi_A$ on scales smaller than some cutoff
scale $\varepsilon$ in an appropriate local frame.  If a low-energy
observer at point $(U_0,V_0)$ uses a test function $f$, then a
low-energy observer at point $x_1$ uses a test function which is
obtained
from the test function $f$ by boosting along the timelike
Killing
vector.  Using the notation of equation \newpoint , this function is
\eqn\lefdeftwo{
f(x-x_1) = f\bigl(e^{t/4GM}(U - U_1), e^{-t/4GM}(V - V_1),
\xv - \xv_1\bigr) \>.
}

The assumption of locality of the nice-slice
theory can be tested by calculating the commutators of low-energy
nice-slice fields
obtained from light-front open bosonic string field theory.  For
technical reasons, we are unable to carry out such a calculation in
the Schwarzschild geometry.  In the limit of large black hole mass,
however, the spacetime region of interest is well approximated by
flat Minkowski space, where the computations are straightforward.

Using light-front coordinates
\eqn\lfc{
x^{\pm} = {1 \over {\sqrt{2}}} \left ( x^0 \mp x^{D-1} \right ) \>,
}
in which the Minkowski line element is
\eqn\Mle{ds^2 = -2d\xp d\xm + \delta_{ij} dx^i dx^j \>, \qquad i,j
\in \{ 1, \ldots, D-2 \} \>,
}
the light-front time $\xp$ corresponds to $U$, the longitudinal
direction $\xm$ corresponds to $V$, and the horizon becomes the
planar lightlike hypersurface $\xp = 0$.  A set of nice slices can be
constructed in Minkowski spacetime in exactly the same way as the
black hole spacetime, except that we must replace asymptotic time $t$
by the Rindler time $\omega = t/4GM$.

In Appendix A we calculate the matrix element
\eqn\goob{
M(1,2;3) = \vev{0|[\Phi(1), \Phi(2)]|3} \>,
}
to leading order in the string coupling $g$.  It gives
the overlap of the commutator of two light-front open bosonic string
fields with state $|3\rangle$.
To leading order in $g$, the commutator creates a single string
state.
The matrix element
\eqn\goobtwo{
M_{AB}(1,2;3) = \vev{0|[\phi_A(x_1), \phi_B(x_2)]|3} \>,
}
for the commutator of any two mass eigenstate fields $\phi_A(x)$ can
be obtained from equation \goob\ by folding in with the appropriate
transverse string wave functions.
We show that there exist matrix elements of this form which are
non-zero
even when $x_1-x_2$ is spacelike.
Finally we obtain, matrix elements
of the commutator of low-energy nice-slice fields
\eqn\goobthree{
\hat M_{AB}(1,2;3)= \vev{0|[{\hat \phi}_A(x_1),{\hat \phi}_B(x_2)]|3}
\>,
}
by integrating the mass eigenstate fields against test
functions appropriate for nice-slice observers at $x_1$ and $x_2$.

Because the spacelike separation between $x_1$ and $x_2$ grows
exponentially with $\omega$, one would ordinarily expect the matrix
elements \goobthree\ to tend to zero very rapidly with $\omega$.
Indeed, for the commutator of tachyon component fields with a
spectator tachyon, this is what we find.  This does
not give a good indication of the ``size'' of the commutator,
however,
because for large $\omega$ the commutator creates a high-mass string
state which has very little overlap with any fixed-mass spectator
state. A better way to obtain the magnitude
of the commutator is to multiply the matrix element \goobthree\ by
its complex conjugate and sum over spectator states;
in other words, to calculate a matrix element of the form
\eqn\goobfour{
\hat M(1,2; 1', 2')= \vev{0|[{\hat \phi}_1(x_1),{\hat
\phi}_2(x_2)][{\hat
\phi}_2({x_2}'),{\hat \phi}_1({x_1}')]|0} \>.
}
A detailed calculation of these matrix elements to leading order in
$g$ is presented in Appendix~B, but we shall describe the basic
points of the calculation here.  The matrix element \goobfour\ has
the form of a sum of finite time scattering amplitudes, which can be
calculated using the first quantized formalism.  The discussion is
simplified if we Fourier transform the amplitude to momentum space.
Since the spectator states are all single string states,
if we define  $s = -(p_1 + p_2)^2$ and $t = -(p_2 - {p_2}')^2$,
then the relevant string world sheet describes an $s$-channel
diagram.
One finds that as $\omega$ increases, the dominant contribution to
the matrix element
comes from the region where the longitudinal momenta $\pp_1$ and
$\pp_{1'}$ are of order $e^{-\omega}$, and the $s$ variable
is large and positive.  Heuristically, what happens is the
scattering amplitudes are driven into the Regge region, where they
behave as $s^{\alpha(t)}$.  In open string theory,
$\alpha(t) =
\alpha' t + 1$ is the leading Regge trajectory.  Because of the Regge
behavior, matrix elements of the form \goobfour\ do
not decrease as $\omega \rightarrow \infty$, but
stay approximately constant.
Moreover, since the
behavior is dominated
by the leading Regge intercept, the extension to closed bosonic
strings is straightforward.  The matrix element
for tachyon component fields in the closed string theory
actually grows like $e^{\omega}$.

One might have argued that the behavior found above should be
expected, since the spectrum of the bosonic string contains a
tachyon.  It should be clear from the above discussion, however, that
the existence of the tachyon has no bearing on the result.  In fact,
if one canonically quantizes a tachyon field in Minkowski space by
imposing that the field commute with itself on some spacelike
surface,
the Lorentz transformation properties of scalar fields guarantee that
the field will commute with itself on all spacelike surfaces.  It is
the
Regge behavior of strings, or in other words the existence of the
infinite tower of massive states to which the commutator can couple,
which governs the behavior of the matrix element \goobfour\ as
$\omega \rightarrow \infty$.  This behavior will therefore also be
present in the tachyon-free superstring case.

The above results have important consequences for the nice-slice
theory.  The effective fields ${\hat \phi}_A(x_1)$ and ${\hat
\phi}_B(x_2)$ are, by construction, low-energy fields as measured by
nice-slice observers.  Since these fields belong to the algebra of
operators of the regulated nice-slice theory, then so must their
commutator.  Our explicit computation shows, however, that the
commutator (to leading order in the string coupling $g$) is an
operator which creates a single excited string which is stretched
between $x_1$ and $x_2$.  The mass squared of this string state is of
order
$s \sim e^{\omega}/\alpha'$, and eventually becomes
trans-Planckian as $\omega$ gets large.  Nevertheless, the nice-slice
energy of the configuration remains low, and this state must
therefore be
regarded as an unavoidable part of the nice-slice theory, which could
never have been discovered if we had first truncated the theory to
the massless fields.

Let us consider how it is possible for configurations to have low
values of the nice-slice energy, and yet have huge trans-Planckian
masses.  Consider a state with a massless particle of low nice-slice
energy at each of the points $x_1$ and $x_2$.  Let the four-momentum
of the particle at $x_2$ be
\eqn\qtwo{
q_2 = (\qm_2, \qp_2, \qv_2 ) \>,
}
with the individual components of $q_2$ being smaller than some
cutoff
momentum.
The four-momentum of a low nice-slice energy particle at $x_1$ is
given by
\eqn\qone{
q_1 = (e^{\omega}\qm_1, e^{-\omega} \qp_1, \qv_1 ) \>,
}
with the components $q_1^{\mu}$ being smaller than the cutoff
momentum.  Although the nice-slice energy of this configuration is
small, the Minkowski mass of the state rapidly becomes
trans-Planckian as $\omega$ increases:
\eqn\masssq{
m^2 = -(q_1 + q_2)^2 \sim e^{\omega} \qm_1 \qp_2
}
Generally, a state of low nice-slice energy will have very large mass
squared if it is spread over a large distance.

It is disturbing at first sight to find not only nonvanishing, but
large values for the commutator of fields with spacelike separated
arguments. Note, however, that this effect is only encountered under
very extreme kinematic circumstances.
In Appendix A.2,
it is shown that if one considers component fields appropriate for
low-energy {\it Minkowski} observers, then the matrix elements
\goobthree\
are given by the usual field-theoretic formulas, and are therefore
suppressed when the fields are spacelike separated.  This is simply
because the
invariant mass squared of this process is always much smaller than
$1/\alpha'$, so the commutator cannot couple to the higher mass,
extended string states.


\newsec{Discussion and Conclusions}

Having obtained the commutators of nice-slice fields from string
field theory, we are now in a position to understand their
significance for the nice-slice argument.  In this section we will
argue that the assumption of locality of the low-energy nice-slice
theory is not valid in string theory.  Evidence for this will be
obtained in two ways.  First, we compare the commutators of nice
slice fields derived from string field theory to their field theory
counterparts, and find that the commutators in string theory exhibit
far more nonlocality.  Second, we argue that mathematical consistency
of the low-energy theory necessitates the inclusion of extremely
nonlocal states corresponding to very massive strings.

\subsec{Comparing Commutators}

Let us begin our comparison by examining the commutators of local
fields more closely.  For the moment, let us ignore gravity, and
focus on quantum field theory on a fixed background spacetime.  The
statement of causality is that no local physical signal can propagate
faster than light.  For a theory of quantum fields on a
(well-behaved) manifold with a fixed metric, there exists a well
defined light cone for each point, and causality is implemented by
requiring the {\it gauge invariant} local fields to commute when
their
arguments are spacelike related.  There is no such requirement for
gauge variant fields.  For example, it is shown in
Appendix A.2 that in Yang-Mills theory the commutator
$[A_i^a(x_1),A_j^b(x_2)]$ of two transverse vector fields in
light-front gauge fails to vanish when $x_1-x_2$ is spacelike.  There
exist
gauge invariant local fields, however, such as $\tr [F^2]$,
and these fields do commute at spacelike separation.

When gravity is included, the situation is more tricky.  Since
the symmetry group of the theory includes diffeomorphisms, there
simply are no local invariants, and one cannot place any restrictions
on the commutators of strictly local fields.  This can be said
another way.
In a quantum theory of gravity, it is impossible to say {\it a
priori} whether two points $x$ and $y$ are spacelike related, so one
cannot impose the condition $[\phi(x),\phi(y)]=0$ as an operator
equation.  One can only calculate the matrix elements of the
commutator in a state of the gravitational field.

Now let us return to string theory and the nice slice.  The
nice-slice
argument assumes that gravitational fluctuations can be
neglected.  Since gravity enters open string theory only at one-loop
level, the calculations of commutators of nice-slice fields
obtained from open string field theory can be directly compared to
their counterparts obtained from quantum field theory in a fixed
background.

Consider first the commutator of two uncharged tachyon fields.
In ordinary quantum field theory, the transformation properties of
scalar fields (tachyonic or not) guarantee that if a scalar field
commutes with itself on one spacelike surface, then it commutes with
itself on all spacelike surfaces.  Open string field theory, on the
other hand, gives a nonvanishing commutator.  Moreover, its
magnitude stays roughly constant in time on the nice slice.
We have thus found an example where gauge invariant fields
fail to commute at spacelike separation in string theory.

The comparison of the commutator of two nice-slice vector fields is
more subtle.  Direct computation shows that the magnitude of this
commutator stays roughly constant in both quantum field theory and in
string theory.  In Yang-Mills theory in light-front gauge, the matrix
element \goobfour\ remains roughly constant because the $A^-$
component of the vector field is a nonlocal functional of the
transverse components.  This nonlocality enters the commutator of the
transverse components at order $g$.  As was mentioned previously, no
significance is attached to this, because there exist gauge invariant
functionals of the gauge fields for which the commutator vanishes.

In string theory, the matrix element \goobfour\ for non-Abelian
vector fields contains the expression obtained in Yang-Mills theory
(as it must), but also contains additional terms.  These extra terms
also remain roughly constant, but do so because of the Regge behavior
of strings, not because of the nonlocality introduced by the
longitudinal components of the gauge fields.  In this case, then, the
most we can say is that string theory introduces additional
nonlocality which arises for entirely different reasons.

In closed string theory, even more extreme effects occur.  For
example, the analog of the matrix element \goobfour\ for closed
string
tachyonic fields increases exponentially in
time.  Note that this additional degree of nonlocality is due to the
shift in the intercept of the leading Regge trajectory from 1 to
2. In other words, it is due to the presence of the graviton.

Finally, one should compare the above findings to those obtained
from string S-matrix calculations
\ref\dalowe{D.~A.~Lowe, {\it The Planckian Conspiracy: String Theory
and the Black Hole Information Paradox,} preprint UCSBTH-95-11,
hep-th/9505074.}.
There one finds a degree of nonlocality far smaller than that present
in the light-front string field commutator.
Although certain S-matrix elements computed in \dalowe\ do
display nonlocal effects over macroscopic separations, these
amplitudes were found to be highly suppressed.
This is to be expected, otherwise an observer crossing an event
horizon would necessarily experience a large scale violation of the
equivalence principle. It is an important open question whether an
interpolating field could be constructed which is ``more local'' than
the
light-front string fields.

\subsec{The Structure of the Nice Slice Theory}

The comparison of the commutators of low-energy nice-slice fields in
quantum field theory and in string theory has provided evidence that
the assumption of locality of the nice-slice theory is not valid.
Now, let us turn our attention to the mathematical structure of the
low-energy nice-slice theory.
The assumption of locality in the nice-slice
argument is an assumption about the result of truncating a
high-energy theory, which includes gravity, to the low nice-slice
energy
modes.  This involves an order of operations.  The order of
operations that is implicit in presenting the nice-slice argument is
to first truncate the high-energy theory down to a system of low-mass
fields, and then to write a theory of these fields on the nice slice.

The appropriate order of operations, however, is to first write down
the
high-energy theory on the nice slice, and then to truncate the system
to the low nice-slice energy modes.  Let us postulate string field
theory as the high-energy theory, and consider the structure of the
low-energy nice-slice theory derived from it.  The set of low-energy
nice-slice degrees of freedom certainly contains the set of fields
obtained by smearing the low-mass component fields of the string
field theory with appropriate test functions.  Let us see what else
might enter the theory.  One requirement we must impose is that the
operator algebra of the low-energy theory is closed under
commutation.  Thus, consider the commutator of two tachyon fields.
In open string field theory,
the magnitude of this commutator \goobfour\ remains roughly constant
in time,
while in closed string field theory it grows like $e^{\omega}$.
The commutator couples strongly to the intermediate states
that dominate the amplitude, and these states must be
included in the low-energy nice-slice theory.  Moreover, the Regge
behavior of the amplitude \goobfour\ shows that these states consist
of
single strings stretched between $x_1$ and $x_2$.  The mass squared
of these states grows like $e^{\omega}$, and becomes trans-Planckian
as $\omega$ increases.  Nevertheless, the nice-slice energy of these
states is low, so they cannot be excluded.  The conclusion is that
the
low-energy nice-slice theory derived from string field theory must
contain more than the usual low-mass fields.  It must also contain
highly nonlocal states of extremely massive strings stretched over
macroscopic distances.

The assumption of approximate locality of the low-energy nice-slice
theory is thus seen to be violated in string theory.  The Hilbert
space of the
theory cannot be factorized into a product of a space of states
inside the horizon with a space of states outside the horizon, and
the argument for information loss breaks down.

It should be noted, of course, that we have in no way proved the
conjecture of black hole complementarity.  We have simply showed that
the usual argument for information loss is no argument at all in
light-front string theory. It is expected that information only
begins to
come out of an evaporating black hole when the black hole area has
reduced to half
its original value, at a time of order $G^2 M^3$ \refs{\page}.  Our
perturbative calculation, however, is only valid for timescales of
order $GM \log (g)$ \refs{\dalowe,\holo}.  If we were to go to higher
order in perturbation theory, the singularity would begin to
manifest itself after a time of order $GM \log (GM^2)$, and at
present we do not have a theory which allows us to deal with this
problem.

Given the results obtained here, it is difficult to imagine how
string theory could be formulated in terms of a $(D-1)-$dimensional
set of local degrees of freedom, at least without incorporating an
enormous amount of gauge symmetry.  Further evidence for this view
has been given previously by numerous authors \refs{\holo, \atwit,
\klebsus,
\gross}.

\bigskip

\noindent
\underbar{Acknowledgements:}
\hfill\break
This work was supported in part by NSF Grant Nos. PHY-91-16964,
PHY-94-07194, and PHY-89-17438.  J.~U. is supported in part by
an NSF Graduate Fellowship.

\appendix{A}{The Commutator of Open Bosonic String Fields}

\subsec{Calculation of the Commutator}

We employ the formalism of light-front open bosonic string field
theory~\refs{\mand,\kaki,\crge,\htc}, and work in D=26-dimensional
Minkowski spacetime with light-front coordinates
\eqn\lccoord{
x^\pm = {1\over \sqrt{2}}(x^0 \mp x^{D-1}) \>.
}
The coordinate $\xp$ is the time coordinate, and in the
light-front gauge we fix the string coordinate $X^+ = \xp$.  The
$D-2$
transverse string coordinates are expanded as
\eqn\strcoord{
\Xv (\sigma) = \xv + 2 \sum_{\ell = 1}^{\infty} \xv_{\ell} \cos
(\ell  \sigma) \>.
}

We are interested in the following commutator
$\left [ \Phi_H(1),\Phi_H(2) \right ]$,
where $\Phi_H(i)=\Phi_H(x^+_i,\xm_i,\Xv_i(\sigma))$ are
light-front open bosonic string field operators in the Heisenberg
picture. We will suppress gauge indices for the most part in the
following.
In the free theory, the commutator of mass
eigenstate
fields vanishes at spacelike separation, which is the kinematical
situation we are interested in~\refs{\martinec, \lowe}.  When string
interactions are included, however, the commutator no longer
vanishes.  This was established in~\refs{\lsu} for fields that are
spacelike separated in the transverse direction but lie on the same
light-front time slice.  In the following, this result is generalized
to spacelike separated fields on different light-front time slices by
evaluating matrix elements of the commutator to leading nontrivial
order in a perturbative expansion in the string coupling $g$.

The light-front Hamiltonian can be written as $H = H_0 +
\sum_{i=1}^{\infty} g^i V_i$, where $H_0$ is the Hamiltonian for
noninteracting string fields, and the leading order interaction term
$V_1$ is the standard cubic string coupling.  In the interaction
picture, the string fields have the expansion
$\Phi_I = \Phi_{\rm a} + \Phi_{\rm c}$ with
\eqn\ipsf{
\Phi_{\rm a} (x^+, \xm, \Xv(\sigma))
= \int {{d^{D-2}p} \over {(2\pi)^{D-2}}} \int_0^{\infty} {{d\pp}
\over {4 \pi \pp}} \sum_{ \{\nv_l \} } e^{ip \cdot x} f_{ \{
\nv_l \} }(
\xv_l) A(\pp, \pv, \{ \nv_l \} )  \> }
where $p \cdot x = -\pmin x^+ - \pp \xm + \pv \cdot \xv$,
and $\Phi_{\rm c} = \Phi_{\rm a}^\dagger$.
The
light-front energy of a string state is given by
\eqn\lcenergy{
\pmin \bigl(\pp, \pv, \{ \nv_l \}\bigr) = {{\pv^{\; 2} + 2\sum_{l, i}
l n_l^i + m_0^2} \over {2\pp}} \>,
}
and the $f_{ \{ \nv_l \} }( \xv_l)$ are wave functions for the modes
of transverse oscillation of the string.  In our conventions,
$\alpha' = \half$.  The mode operators obey the canonical commutation
relations
\eqn\commut{
\eqalign{
\Bigl [ A(\pp, \pv, \{ \nv_l \} ), &A^{\dag}( {\pp}', {\pv}\,', \{
\nv_l \,' \} ) \Bigr ] \cr
&= 2 \pp (2\pi)^{D-1} \delta(\pp-{\pp}')
\delta^{D-2}(\pv - {\pv}\,') \delta_{ \{ \nv_l \}, \{ \nv_l\,' \} }
\>. \cr
}
}

Consider the matrix element
\eqn\Mdef{
M(1,2;3) = \vev{0|[\Phi_H(1),\Phi_H(2)]|3} \>,
}
where $|0\rangle$ is the vacuum state and $|3\rangle$ is a spectator
state, which is necessary in order to have a nonvanishing
contribution at first order in the string interaction.  Written in
the interaction picture, the matrix element \Mdef\ becomes
\eqn\Mint{
M(1,2;3) = {}_I \vev{0|\Phi_I(1)U_I(x^+_1,x^+_2)\Phi_I(2)
|3;x^+_2}_I - (1
\leftrightarrow 2)
}
where $U_I$ is the interaction picture time evolution operator.
Using the Feynman-Dyson expansion for $U_I$, $M$ can be expanded in
powers of the string coupling $g$ as $M = \sum_{i=0}^{\infty} g^i
M^{(i)}$.  The zeroth order term is simply a matrix element of the
commutator of two interaction picture string fields,
\eqn\Mzero{
M^{(0)}(1,2;3) =  \vev{0|[\Phi_I(1),\Phi_I(2)]|3} \>,
}
which vanishes for the kinematics we are interested
in~\refs{\martinec, \lowe}.

Consider the next term in the expansion for $M$.  It will prove
convenient to choose $|3\rangle$ to be an eigenstate of $H_0$.
After some algebra, the first order term can be written
\eqn\Minty{
\eqalign{
M^{(1)}(1,2;3) = i g \int_{x^+_1}^{x^+_2} d\xp \vev{0 | \{ &\Phi_I
(1)
V_1 (\xp) \Phi_I (2) + \Phi_I (2) V_1 (\xp) \Phi_I (1) \cr
&- \Phi_I (1) \Phi_I (2) V_1 (\xp)\} | 3 } \>. \cr
}
}
Separating the vertex into terms with given numbers of creation and
annihilation operators, $V_1 = V_{1\rm aaa} + V_{1\rm aac} + V_{1\rm
acc}
+ V_{1\rm ccc}$, one finds that the matrix element is
nonvanishing only when the spectator is a single string state,
\eqn\specone{
| 3 \rangle = A^{\dagger} (\pp_3, \pv_3, \{ \nv_{\ell, 3} \} ) | 0
\rangle \>,
}
and the only terms which contribute are
\eqn\ancre{
\vev{0 | \{ \Phi_{\rm a} (1)
V_{1\rm aac} (\xp) \Phi_{\rm c} (2) + \Phi_{\rm a} (2) V_{1\rm aac}
(\xp)
\Phi_{\rm c} (1) - \Phi_{\rm a} (1) \Phi_{\rm a} (2) V_{1\rm acc}
(\xp)
\} | 3 }.
}
The leading order contribution
to the matrix element \Minty\ with a single string spectator is
given by
\eqn\Mresult{
\eqalign{
M^{(1)}(1&,2;3)
= (2\pi)^{D-1} g \left ( \prod_{r=1}^2 \int {{d^{D-2} p_r}
\over {(2\pi)^{D-2}}} \int_{-\infty}^{\infty} {{d\alpha_r} \over
{4\pi | \alpha_r |}} \sum_{ \{ \nv_{\ell, r} \} } f_{ \{ \nv_{\ell,
r} \} } ( \{ \xv_{\ell, r} \} ) \right ) \cr
&\times F(\alpha_1, \alpha_2 ) \>
\Bigl( \sum_{r=1}^3 \pmin_r \Bigr)^{-1}
\left [
\exp \Bigl( -i x^+_1 \sum_{r=1}^3 \pmin_r \Bigr) - \exp \Bigl(
-i x^+_2 \sum_{r=1}^3 \pmin_r \Bigr) \right ] \cr
&\times
\exp \Bigl( i \sum_{r=1}^2 p_r \cdot x_r \Bigr)
\delta \Bigl( \sum_{r=1}^3
\alpha_r \Bigr) \delta^{D-2} \Bigl( \sum_{r=1}^3 \pv_r \Bigr)
\left [  {\tilde V}_1 (1;2;3) +
{\tilde V}_1 (2;1;3) \right ] \>, \cr
\cr
}
}
where ${\tilde V}_1$ is the 3-string vertex in momentum space
and in the occupation number basis $\{ \nv_{\ell, r} \}$.
In the above, we have defined $| \alpha_r | = 2\pp_r$ and $\pmin_r =
(\pv_r^{\> 2} + m^2_r ) / \alpha_r$.  The sign of $\alpha_r$ is
positive (negative) for incoming (outgoing) strings.  We have also
defined
\eqn\Fdef{
F(\alpha_1, \alpha_2) = \Theta ( \alpha_1 ) \Theta ( -\alpha_2 ) +
\Theta ( -\alpha_1 ) \Theta ( \alpha_2 ) - \Theta ( -\alpha_1 )
\Theta ( -\alpha_2 ) \>,
}
where $\Theta$ is the
Heaviside function, the three terms corresponding to the three terms
in the matrix element \ancre.

The first question we want to address is whether this matrix element
vanishes when the string fields in the commutator are at spacelike
separation, as it would, for example, if we were dealing with a local
field theory of scalar fields.  For this purpose, let us specialize
to the case of tachyon component fields and a tachyon spectator
state.
This simplifies the analysis considerably, as the momentum space
representation of the three tachyon vertex reduces to
\eqn\Tvert{
{\tilde V}_1
\left ( \alpha_1, \pv_1; \alpha_2, \pv_2; \alpha_3, \pv_3 \right )
= \exp \Bigl( {{\tau_0} \over 2} \sum_{r=1}^3 \pmin_r \Bigr) \>,
}
where $\tau_0 = \sum_{r=1}^3 \alpha_r \log | \alpha_r |$.
We can then write
\eqn\Mtach{
\eqalign{
\vev{ 0 | [ T(x_1), T(x_2)& ] | 3} = 2(2\pi)^{D-1} g \left (
\prod_{r=1}^2 \int {{d^{D-2} p_r} \over {(2\pi)^{D-2}}}
\int_{-\infty}^{\infty} {{d\alpha_r} \over {4\pi | \alpha_r |}}
\right )\, F(\alpha_1, \alpha_2 ) \cr
&\times \Bigl( \sum_{r=1}^3 \pmin_r \Bigr)^{-1}
\left [\exp \Bigl( -i x^+_1 \sum_{r=1}^3 \pmin_r \Bigr)
- \exp \Bigl( -i x^+_2 \sum_{r=1}^3 \pmin_r \Bigr) \right ] \cr
&\times \delta \Bigl( \sum_{r=1}^3
\alpha_r \Bigr) \delta^{D-2} \Bigl( \sum_{r=1}^3 \pv_r \Bigr)
\exp \Bigl( {{\tau_0} \over 2} \sum_{r=1}^3 \pmin_r \Bigr)
\> e^{ip_1 \cdot x_1 +ip_2 \cdot x_2}\, . \cr
}
}

If the cubic string vertex \Tvert\ were polynomial in the
longitudinal momenta~$\alpha_r$, the matrix element \Mtach\ would
vanish
for
$x_1-x_2$ spacelike, by the usual contour deformation argument
\refs{\bd}.
The vertex is not polynomial, however, so the usual cancellation
between terms fails, leaving behind a nonvanishing answer for the
matrix element.

This is an important sign that the nice-slice argument may fail
in light-front string theory.  We must, however, do more work to
establish this result.  On the one hand, we need to show that the
commutator is significantly different from zero for spacelike
separated fields on a nice slice, while at the same time this
effect should be very much suppressed under the kinematic conditions
found in everyday experiments at low energies.  These issues will
each be addressed in what follows.

\subsec{Correspondence with Low Energy Minkowski Field Theory}

Local quantum field theory in Minkowski space provides a very good
description of the low-energy world we observe, and any unified
theory should reproduce the structure of local Minkowski field theory
for kinematic situations appropriate to low-energy Minkowski
observers.  This structure includes the requirement that the
commutator of gauge-invariant local fields must vanish at spacelike
separation.  In this section, we show that the low-energy fields
obtained from light-front string field theory satisfy this
requirement.

Consider a Minkowski observer, whose measuring apparatus is sensitive
to frequencies of order $E \ll 1 / \sqrt{\alpha'}$.  Such an observer
will describe physics by a set of low-energy fields ${\hat \phi}_A
(x)$, which can be obtained from the component fields $\phi_A$ of
light-front string field theory by integrating $\phi_A(y)$ against a
test function $f(x-y)$.  The Fourier components ${\tilde f}(q)$ of
$f$ have support only for $q^{\mu} {\
\lower-1.2pt\vbox{\hbox{\rlap{$<$}\lower5pt\vbox{\hbox{$\sim$}}}}\ }
E$.

Let us consider the vector field, which clearly enters the low-energy
Minkowski theory.  The interaction picture vector field is expanded
as
\eqn\Aexp{
A_{\mu}^b (x) = \sum_{\lambda=1}^{24} \int {{d^{D-2} p} \over
{(2\pi)^{D-2}}} \int {{d \pp} \over {4\pi \pp}} \left [
\epsilon_{\mu}(\lambda) \> a(\pp, \pv, \lambda, b) \> e^{ip \cdot x}
+
\hbox{h. c.} \right ] \>,
}
where the $\epsilon_{\mu}(\lambda)$ are polarization vectors which
correspond to the polarization states $\lambda$ and $b$ is the
relevant group index.  In the light-front
gauge, $A^+ = 0$, and $A^-$ is expressed as a nonlocal function of
the $D-2$ transverse components $A^i$.  The commutation relations for
$A^-$ will therefore be nonlocal even in free field theory.  For this
reason, we restrict ourselves to commutators of the transverse
fields, which do vanish in free field theory.  The matrix element
involving massless vectors can now be
calculated using equation \Mresult , and we obtain
\eqn\Mlowen{
\eqalign{
&\vev{0| [{\hat A}^a_i(x_1),{\hat A}^b_j(x_2)] |3,c} = (2\pi)^{D-1} g
f^{abc} \left ( \prod_{r=1}^2 \int {{d^{D-2} p_r}
\over {(2\pi)^{D-2}}} \int_{-\infty}^{\infty} {{d\alpha_r} \over
{4\pi | \alpha_r |}} \right ) \cr
&\times {{F(\alpha_1, \alpha_2 )} \over {\sum_{r=1}^3 \pmin_r }} \>
\delta \Bigl( \sum_{r=1}^3 \alpha_r \Bigr)
\delta^{D-2} \Bigl( \sum_{r=1}^3 \pv_r \Bigr) {\tilde V}_1
(1,i;2,j;3) \> e^{ ip_1 \cdot x_1 +ip_2 \cdot x_2 } \cr
&\times \left [ {\tilde f}_1
(-p_2 - p_3 ) {\tilde f}_2 (p_2) e^{ -i x^+_1 \sum_{r=1}^3 \pmin_r }
- {\tilde f}_1 (p_1) {\tilde f}_2 (-p_1 - p_3 ) e^{ -i x^+_2
\sum_{r=1}^3 \pmin_r } \right ] \>. \cr
}
}
Consider the
first term in the square brackets.  The function ${\tilde f}_2$
constrains $p^{\mu}_2 \sim E$, and the function ${\tilde f}_1$
constrains the sum $p^{\mu}_2 + p^{\mu}_3 \sim E$, which
implies that $p^{\mu}_3 \sim E$ as well.  A similar argument
holds for the second term.  In other words, the commutator of two
low-energy operators is itself a low-energy operator, and does not
couple
to states of high energy.  Therefore, for the matrix element \Mlowen\
to be non-negligible, the state $|3,c\rangle$ must be a
low-energy state, which we will select to be a vector boson polarized
along the
k direction.  The three vector boson vertex is
\eqn\Avert{
{\tilde V}_1 (1,i;2,j;3,k) = \left [ \delta_{ij} {{{\cal P}_k} \over
{\alpha_3}} + \delta_{jk} {{{\cal P}_i} \over {\alpha_1}}  +
\delta_{ki} {{{\cal P}_j} \over {\alpha_2}} + 2\alpha' {{{\cal P}_i
{\cal P}_j {\cal P}_k} \over {\alpha_1 \alpha_2 \alpha_3}} \right ]
\exp \Bigl( {{\tau_0} \over 2} \sum_{r=3}^3 \pmin_r \Bigr)
\>,
}
where ${\vec {\cal P}} = \alpha_1 \pv_2 - \alpha_2 \pv_1$ is
cyclically symmetric.  We can now substitute this expression into
equation \Mlowen .

Since the test functions restrict the allowed momenta to be of order
$E \ll 1/\sqrt{\alpha'}$, we keep only the leading terms in
$\alpha'$.  The only dependence on $\alpha'$ comes from the vertex
\Avert , both from the explicit dependence shown and through
$\tau_0$, which is properly written
\eqn\adep{
\tau_0 = \alpha' \sum_{r=3}^3 \alpha_r \log (\alpha' \alpha_r^2 )
\>.
}
Dropping the test functions, the leading term of the commutator is
\eqn\Acomm{
\eqalign{
\vev{ 0 | [ {\hat A}^a_i&(x_1), {\hat A}^b_j(x_2) ] |3,c,k} =
(2\pi)^{D-1} g f^{abc} \left ( \prod_{r=1}^2 \int {{d^{D-2} p_r}
\over {(2\pi)^{D-2}}} \int_{-\infty}^{\infty} {{d\alpha_r} \over
{4\pi | \alpha_r |}} \right ) \cr
&\times {F(\alpha_1, \alpha_2 ) \over \sum_{r=1}^3 \pmin_r } \>
\delta \Bigl( \sum_{r=1}^3 \alpha_r \Bigr) \delta^{D-2} \Bigl(
\sum_{r=1}^3 \pv_r \Bigr) e^{ ip_1 \cdot x_1 +ip_2 \cdot x_2} \cr
&\times \left [ \delta_{ij} {{{\cal P}_k} \over {\alpha_3}} +
\delta_{jk} {{{\cal P}_i} \over {\alpha_1}}  + \delta_{ki} {{{\cal
P}_j} \over {\alpha_2}} \right ] \left [ e^{-i x^+_1 \sum_{r=1}^3
\pmin_r} - e^{-i x^+_2 \sum_{r=1}^3 \pmin_r} \right ] \>.
}
}
which is the same result as that obtained from Yang-Mills theory in
the light-front gauge.

The commutator \Acomm\ does not vanish for $x_1-x_2$ spacelike,
because the integrand is nonpolynomial in the longitudinal momenta
$\alpha_r$.  This does not
violate the rules of local field theory, however, because the fields
$A^a_i$ are not gauge invariant.  It is only the commutators of
gauge invariant local fields, such as $\tr F^2$, which are
required to vanish when the arguments of the fields are spacelike
related.  This ensures that a gauge invariant signal cannot propagate
faster than the speed of light.  What we have shown is that the
commutator of low-energy gauge fields obtained from light-front
string field theory is exactly the same as that predicted by
low-energy field theory.  In other words, the commutator of low
Minkowski
energy fields does not acquire any {\it additional} nonlocality in
string theory.  Therefore, one may construct the usual gauge
invariant fields from the $A^a_{\mu}$, and they will commute at
spacelike separations.

Light-front string field theory has thus passed an important test.
Had we found that light-front string field theory introduces
additional nonlocality into the commutator of low-energy fields, it
would have indicated a catastrophic breakdown in the low-energy
predictions of the theory, and the commutator (and perhaps the theory
itself) could not be taken seriously.  As it stands, the theory
avoids introducing any additional nonlocality at low energy, and we
can proceed to study the predictions it makes for higher energy.

\subsec{The Tachyon Commutator on a Nice Slice}

In equation \Mtach , we considered the overlap of the state
$[T(x_1),T(x_2)]|0\rangle$ with a single tachyon state.  We would now
like to consider the behavior of this matrix element when the fields
$T(x_1)$ and $T(x_2)$ are replaced with low-energy nice-slice fields
${\hat T}(x_1)$ and ${\hat T}(x_2)$.  Following an analysis similar
to that in Subsection A.2, it is easy to see that the test functions
defined by equations \lefdef\ and \lefdeftwo\
restrict $\pmin_2 \sim E$, as before, but now $\pmin_3 \sim
{\alpha'}^{-1/2} e^{\omega}$.  The momentum conservation
delta-functions restrict $\alpha_3$ and $p_3^i$ to be of order $E$,
so the only way to make $\pmin_3$ large is to have a very large mass
squared.  Therefore, the three tachyon amplitude goes to zero quickly
as $\omega$ increases.

This does not mean that the commutator of two tachyons is ``small''
but simply that its overlap with single tachyon states is.  This is
not surprising in view of our expectation that the commutator
creates a high-mass extended string state from the vacuum.
The aim of Appendix~A.1 was only to establish that the commutator
of string fields can be nonvanishing at spacelike separation and
we focused on the case of a tachyon spectator for simplicity.  In
Appendix~B we will evaluate matrix elements which involve a sum
over spectator states and should therefore pick up the dominant
overlap with the commutator.

\appendix{B}{The Magnitude of the Commutator}

\subsec{A Four Point Amplitude Involving Commutators}

In this Appendix we present the calculation of the matrix element
\eqn\Cdef{
C(4;3;2;1) = \vev{0|[\Phi_H(4),\Phi_H(3)][\Phi_H(2),\Phi_H(1)]|0} \>.
}
As before, we will use perturbation theory to calculate $C$ to
leading nontrivial order in $g$.  Inserting a complete set of
states $\{ | \gamma \rangle \}$, equation \Cdef\ can be written
\eqn\CandM{
C(4;3;2;1) = \sum_{\gamma} M(4,3;\gamma) M(1,2;\gamma)^* \>,
}
where $M$ is given by equation \Mdef .  The first term which does not
vanish identically when the centers of mass of the string fields are
spacelike related is the second order term
\eqn\Cprime{
C^{(2)}(4;3;2;1) = \sum_{\gamma} M^{(1)}(4,3;\gamma)
M^{(1)}(1,2;\gamma)^* \>.
}
It was stated in Appendix~A that $M^{(1)}(1,2;\gamma)$ vanishes
unless
$|\gamma\rangle$ is a single string state.  Thus we can perform the
sum, using the earlier results~\Minty\ and~\ancre.  This leads to a
sum
of nine terms.
Any one of these terms can be isolated by integrating
the expression against appropriate functions of the longitudinal
momenta, so they cannot cancel one another.  We shall present the
calculation for only one of those terms, the term in which $\alpha_r
> 0$ for all $r$:
\eqn\abone{
\eqalign{
C^{(2)}&(4;3;2;1) = \cr
&g^2 \int_{x^+_4}^{x^+_3} dx'^+ \int_{x^+_1}^{x^+_2} dx''^+ \>
\vev{0 |
\Phi_{\rm a} (4) \Phi_{\rm a} (3) V_{1\rm acc} (x'^+)
V_{1\rm aac} (x''^+) \Phi_{\rm c} (1) \Phi_{\rm c} (2) |0}. \cr
}
}
Transforming to a momentum and occupation number representation
(the measure being as in the first line of~\Mresult), the matrix
element is
is
\eqn\bbone{
\eqalign{
\tilde C^{(2)}&(4;3;2;1) \cr
&= g^2 \int_{x^+_4}^{x^+_3} dx'^+ \int_{x^+_1}^{x^+_2} dx''^+ \>
\vev{0 |
A (4) A (3) V_{1\rm acc} (x'^+)
V_{1\rm aac} (x''^+) A^\dagger (1) A^\dagger (2) |0}. \cr
}
}

The time and operator orderings in the matrix element~\bbone\ differ,
but
we may calculate first the Euclidean time ordered matrix element and
obtain the needed Minkowski result by analytic continuation.  The
fields will be at Euclidean times
$\sigma^0_r$ and the interactions at
$\sigma'^0$, $\sigma''^0$, with
$\sigma^0_{3,4} > \sigma'^0 > \sigma''^0 > \sigma^0_{1,2}$.  The
matrix
element in equation~\bbone\ is then
given by

\ifig\fthree{String diagram for four-point amplitude.}
{\epsfxsize=5.5cm \epsfysize=3.5cm \epsfbox{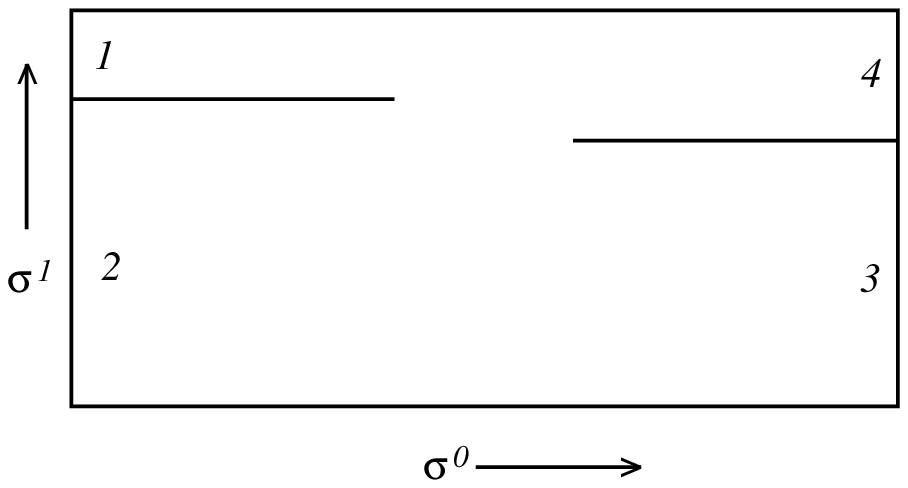}}

\eqn\calp{
\left( \prod_{s = 1}^4 2 \alpha_s^+ \right)^{1/2}
2\pi \delta (\alpha_1 + \alpha_2 - \alpha_3 - \alpha_4)
{\cal P}(1,2,3,4;\sigma'^0,\sigma''^0),
}
where $\cal P$ is the functional integral on the
string world sheet shown in \fthree, (the other cyclic
ordering,~1243,
does  not contribute to the Regge behavior of interest).

It is convenient to take the external times $|\sigma^0_r|$ to be
large.  The
inverse of the mapping
\eqn\mapone{
\rho(z) = \sigma^0 + i\sigma^1 = \alpha_1 \log (z) + \alpha_2 \log
(z-x)
- \alpha_3 \log (1-z) + \bar \rho \>
}
then takes the string world-sheet $\Sigma$ to the
upper half $z$-plane minus four small semicircles surrounding the
points $0,x,1$, and $\infty$.  The parameters $x$ and $\bar\rho$ in
the
mapping are determined implicitly in terms of the interaction times
$\sigma'^0$, $\sigma''^0$ by the condition that
$\partial_z\rho = 0$
at the interaction points,
\eqn\intime{
0  = \frac{\alpha_1}{z} + \frac{\alpha_2}{z - x}
 + \frac{\alpha_3}{1-z} \>.
}
Solving this (quadratic) equation for $z', z''$, the equations
$\rho(z')
= \sigma'^0 + i \pi \alpha_3$, $\rho(z'') = \sigma''^0 + i \pi
\alpha_2$
give two relations between $\sigma'^0$, $\sigma''^0$ and $x$,
$\bar\rho$.
Note that the difference of these relates $x$ to $\delta =
\sigma'^0 - \sigma''^0$.
Strings 1, 2, 3, 4 are mapped to the
semicircles, with respective radii
\eqn\radii{
\eqalign{
r_1 &= x^{-\alpha_2/\alpha_1} e^{(\sigma^0_1 - \bar\rho)/\alpha_1}
\cr
r_2 &= x^{-\alpha_1/\alpha_2} (1 - x)^{\alpha_3/\alpha_2}
  e^{(\sigma^0_2 - \bar\rho)/\alpha_2} \cr
r_3 &= (1 - x)^{\alpha_2/\alpha_3} e^{(\bar\rho -
\sigma^0_3)/\alpha_3} \cr
r_4 &= e^{(\bar\rho - \sigma^0_4)/\alpha_4} \>,
}
}
$r_4$ being the radius in the inverse coordinate $z^{-1}$.

Then
\eqn\conp{
\eqalign{
{\cal P}(1,2,3,4;\sigma'^0,\sigma''^0) &=
J(\alpha_s, \sigma_s^0, \sigma'^0,\sigma''^0)\>
 {\cal
P}'(1,2,3,4;\sigma'^0,\sigma''^0)  \cr
&= J(\alpha_s, \sigma_s^0, \sigma'^0,\sigma''^0) \>
\left\langle \prod_{s=1}^4 r_s^{h_s} {\cal V}_s \right\rangle.
}
}
Here ${\cal P}'$ is the path integral on the upper half-plane minus
semicircles and $J$ is the determinant from the conformal mapping.
In the second line the small semicircles have been replaced by the
corresponding vertex operators.  The semicircles correspond to vertex
operators renormalized at scale $r_s$; the factor
$r_s^{h_s}$ with $h_s$ the weight of ${\cal V}_s$ relates these to
normal ordered vertex operators.  It can be
thought of as arising from the radial evolution from radius $r$ to a
standard radius~1.
To simplify the analysis, we will consider the case of four tachyon
component fields, for which equation \conp\ is
\eqn\contach{
\eqalign{
{\cal P}(1,2,3,4;\sigma'^0,\sigma''^0)
&= J(\alpha_s, \sigma_s^0, \sigma'^0,\sigma''^0)\,
(2\pi)^{D-2} \delta^{D-2} ( \pv_1 + \pv_2 - \pv_3 - \pv_4 ) \cr
&\times x^{\vec p_1 \cdot \vec p_2} (1 - x)^{- \vec p_2 \cdot \vec
p_3}
\prod_{s=1}^4 r_s^{\vec p^{\, 2}_s/2} \>.
}
}
The determinant $J$ can now be determined by
comparison with the on-shell four-point function, using the fact that
$J$
is independent of the transverse momenta,
\eqn\detj{
\eqalign{
J(\alpha_s, \sigma_s^0, \sigma'^0,&\sigma''^0) =
\left( \prod_{s = 1}^4 2 \alpha_s^+ \right)^{-1/2} \frac{dx}{d\delta}
\cr
\times& x^{\alpha_1/\alpha_2 + \alpha_2/\alpha_1}
(1-x)^{-\alpha_3/\alpha_2 - \alpha_2/\alpha_3}
\exp \left\{\sum_{s=1}^4 (- \sigma^0_s + \bar\rho) \theta_s/\alpha_s
\right\} \>,
}
}
with $\theta_s = +1$ for $s=1,2$ and $\theta_s = -1$ for $s= 3,4$.
The first factor cancels the noncovariant factor in the matrix
element~\calp.

The result for the matrix element~\bbone\ is
\eqn\confin{
\eqalign{
\tilde C^{(2)}&(4;3;2;1) = g^2 (2\pi)^{D-1}
\delta (\alpha_1 + \alpha_2 - \alpha_3 - \alpha_4)  \delta^{D-2} (
\pv_1 +
\pv_2 - \pv_3 - \pv_4 )
\cr
&\times \int_{x^+_4}^{x^+_3} dx'^+ \int_{x^+_1}^{x^+_2} dx''^+\>
\frac{dx}{d\delta}
\> x^{p_1 \cdot p_2} \> (1-x)^{-p_2 \cdot p_3}
\exp \left\{\sum_{s=1}^4 (i x^+_s - \bar\rho) p_s^- \theta_s
\right\}.
}
}
The straightforward continuation~$\sigma_s^0 \to i x^+_s$ has been
carried out.  The continuation $\sigma'^0, \sigma''^0 \to ix'^+,
ix''^+$ is
also implicit; this makes $x$ and $\bar\rho$ complex.
The matrix element~\confin\ cannot be evaluated in closed form.
Solving the
conformal
mapping to express all quantities in terms of $x$ and $\sigma'^0$,
the
integral over $\sigma'^0$ can be carried out but the result is a
rather
complicated integral over $x$, with endpoints that are determined
only
implicitly.  However, the important behavior will be determined
simply from
scaling.

\subsec{The Behavior of the Commutator of Nice Slice Fields}

Low-energy nice-slice fields are defined as in Section 3.  As before,
we consider the case of four tachyon component fields.  We treat the
mass squared of the tachyon as a
small positive parameter; we can do this, for example, by giving the
external tachyons momenta in a compactified direction.  This is
ultimately
justified by  considering superstring theory where the calculations
will
yield essentially the same qualitative results.  The commutator in
free
superstring field theory was studied in Ref.~\lowe\ where further
discussion of this issue can be found.

We smear the fields $T(1)$ and $T(4)$ with test functions appropriate
to position $y_1$, and the fields $T(2)$ and $T(3)$ with test
functions appropriate to position $y_2$.  The amplitude under
consideration is thus
\eqn\wutwewant{
{\hat C} = \vev{0|[{\hat T}(y_1),{\hat T}(y_2)][{\hat T}(y_2),{\hat
T}(y_1)]|0} \>.
}
This is the position-space matrix element $C(4;3;2;1)$ folded into
\eqn\testf{
\eqalign{
& f \bigl ( e^{\omega} (x^+_1{-}y^+_1), e^{-\omega} (x^-_1{-}y^-_1),
\xv_1{-}\yv_1 \bigr ) \> f (x_2-y_2) \cr
\times & f (x_3-y_2) \> f \bigl ( e^{\omega} (x^+_4{-}y^+_1),
e^{-\omega}
(x^-_4{-}y^-_1),
\xv_4{-}\yv_1 \bigr ) \>.
}
}
Thus,
\eqn\small{
\alpha_1, \alpha_4, x^+_1, x^+_4 \ \propto\
e^{-\omega}
}
with $\alpha_{2}$, $\alpha_{3}$, $x^+_{2}$,
$x^+_{3}$ and the
transverse quantities approaching constants.  It is straightforward
to read
off the scaling from the result~\confin.  The momenta $p^-_1$ and
$p^-_4$
are large, $O(e^{\omega})$, so the phase factor is highly oscillatory
and
damps the integral unless
\eqn\smtimes{
x'^+ - x_4^+ = O(e^{-\omega}), \qquad x''^+ - x_1^+ = O(e^{-\omega})
}
This in turn implies that $\delta$, $\bar\rho$, and $1-x$ are
$O(e^{-\omega})$, while $dx/d\delta$ approaches a constant.  The
measures
for the convolution and the Fourier transform are covariant and so
constant under the scaling.  The only scaling then comes from the
ranges of
the time integrals and the factor $(1-x)^{-p_2 \cdot p_3}$, with the
result
\eqn\scale{
{\hat C} \ \sim\ e^{\omega(p_2 \cdot p_3 - 2)}\ =\ e^{\omega t/2}
\ =\ e^{\omega(\alpha(t) - 1)}\>
}
in terms of $t = -(p_2 - p_3)^2$ and the tachyon Regge trajectory
$\alpha(t) = \frac{1}{2}t + 1$.
Now $p_2 - p_3$ is in general a spacelike vector, so $t < 0$, but in
all cases $|t| \ll 1/\alpha'=2$, so the magnitude of the commutator
stays essentially constant as $\omega \rightarrow \infty$.  This
behavior is completely different than that exhibited by the
commutator of two local scalar fields, which vanishes when the
fields are spacelike separated.

\subsec{Extension to Closed Bosonic Strings and Superstrings}

Our calculation was only carried out at tree level, in part
because the inconsistency of bosonic string theory prevents a
further analysis.  Since gravity does not appear in the open
bosonic string at tree level, we do not expect to see gravitational
effects in the commutator calculated above.  The closed bosonic
string does contain gravity at tree level, however, so it is of
interest to repeat the calculation for this case.
The
result is easily anticipated: since the amplitude is
dominated by the Regge behavior of the strings, we simply need to
replace the open string Regge trajectory $\alpha(s) = \half s + 1$ by
the closed string trajectory $\alpha(s) = \half s + 2$.  Explicit
calculation of this amplitude shows that this is indeed correct.
For the closed string, then,
the amplitude actually grows like $e^{\omega}$.  Superstrings exhibit
the same behavior, although the calculation is considerably
more complicated.

\subsec{A Limiting Case}

For completeness we evaluate the commutator~\confin\ in a limit where
it
is possible to obtain an analytic expression, namely $\alpha_1,
\alpha_4
\to 0$ with other quantities held fixed.
Solving for the interaction points gives
\eqn\intz{
\eqalign{
&z' = \frac{\alpha_2 (1-x)}{\alpha_4} + O(\alpha_{1,4}^0),
\qquad
z'' = \frac{\alpha_1 x}{\alpha_2 (1-x)} + O(\alpha_{1,4}^2) \cr
&\sigma'^0 = \bar\rho - \alpha_4 \log\frac{\alpha_4}{\alpha_2 e
(1-x)} +
O(\alpha_{1,4}^2) \cr
&\sigma''^0 = \bar\rho + \alpha_2 \log x + \alpha_1
\log\frac{\alpha_1
x}{\alpha_2 e (1-x)} + O(\alpha_{1,4}^2) \cr
&x = e^{-\delta/\alpha_2} + O(\alpha_{1,4}).
}
}
We have kept some second order terms which are needed because some of
the
exponents in the commutator are of order $\alpha_{1,4}^{-1}$.
Inserting the values~\intz\ into the commutator~\confin\ gives
\eqn\conlim{
\eqalign{
\tilde C^{(2)}(4;3;2;1)& = g^2 (2\pi)^{D-1}
\delta (\alpha_1 + \alpha_2 - \alpha_3 - \alpha_4)  \delta^{D-2} (
\pv_1 +
\pv_2 - \pv_3 - \pv_4 )
\cr
&\times e^{2 - (\pv_1^{\, 2} + \pv_4^{\, 2})/2} \alpha_2^{1 -
(\pv_1^{\, 2} +
\pv_4^{\, 2})/2}
\alpha_1^{-1 + \pv_1^{\, 2}/2}  \alpha_4^{-1 + \pv_4^{\, 2}/2} \cr
&\times \int_{x^+_4}^{x^+_3} dx'^+ \int_{x^+_1}^{x^+_2} dx''^+\>
\> \Bigl\{ x^{-1 + (\pv_1 + \pv_2)^2/2} \> (1-x)^{-\pv_1 \cdot \pv_4}
\cr
&\qquad\quad
\times e^{i p_1^-(x_1^+ - x''^+) + i p_2^-(x_2^+ - x''^+) -
i p_3^-(x_3^+ - x'^+) - ip_4^-(x_4^+ - x'^+) } \Bigr\}
}
}
The light-cone energies~$p^-_1$, $p^-_4$ are $O(\alpha_{1,4}^{-1})$
and so
the time integrals are highly oscillatory and are dominated by the
lower
endpoints.  Thus,
\eqn\conlimfin{
\eqalign{
\tilde C^{(2)}(4;3;2;1) &= g^2 (2\pi)^{D-1}
\delta (\alpha_1 + \alpha_2 - \alpha_3 - \alpha_4)  \delta^{D-2} (
\pv_1 +
\pv_2 - \pv_3 - \pv_4 )
\cr
&\times e^{2 - (\pv_1 - \pv_4)^2/2} \alpha_2^{1 - (\pv_1^{\, 2} +
\pv_4^{\, 2})/2}
\alpha_1^{\pv_1^{\, 2}/2}  \alpha_4^{\pv_4^{\, 2}/2} \frac{(i x_4^+ -
i
x_1^+)^{-\pv_1
\cdot \pv_4 }}{(1 - \pv_1^{\> 2}/2)(1 - \pv_4^{\> 2}/2)}\cr
&
\times e^{ i p_2^-(x_2^+ - x_1^+) -
i p_3^-(x_3^+ - x_4^+) }
}
}
Under simultaneous scaling of $\alpha_{1,4}$ and $x_{1,4}^+$ one
recovers
the behavior found previously.
\listrefs
\end